
\input lanlmac
\input tables 



\def\'{\prime}

\def\to{\rightarrow}
\def\hf{{1 \over 2}}
\def\d{\partial}

\def\sitarel#1#2{\mathrel{\mathop{\kern0pt #1}\limits_{#2}}}
\def\inbar{\,\vrule height1.5ex width.4pt depth0pt}
\def\IB{\relax{\rm I\kern-.18em B}}
\def\IC{\relax\hbox{$\inbar\kern-.3em{\rm C}$}}
\def\ID{\relax{\rm I\kern-.18em D}}
\def\IE{\relax{\rm I\kern-.18em E}}
\def\IF{\relax{\rm I\kern-.18em F}}
\def\IG{\relax\hbox{$\inbar\kern-.3em{\rm G}$}}
\def\IH{\relax{\rm I\kern-.18em H}}
\def\II{\relax{\rm I\kern-.18em I}}
\def\IK{\relax{\rm I\kern-.18em K}}
\def\IL{\relax{\rm I\kern-.18em L}}
\def\IM{\relax{\rm I\kern-.18em M}}
\def\IN{\relax{\rm I\kern-.18em N}}
\def\IO{\relax\hbox{$\inbar\kern-.3em{\rm O}$}}
\def\IP{\relax{\rm I\kern-.18em P}}
\def\IQ{\relax\hbox{$\inbar\kern-.3em{\rm Q}$}}
\def\IR{\relax{\rm I\kern-.18em R}}
\font\cmss=cmss10 \font\cmsss=cmss10 at 7pt
\def\IZ{\relax\ifmmode\mathchoice
{\hbox{\cmss Z\kern-.4em Z}}{\hbox{\cmss Z\kern-.4em Z}}
{\lower.9pt\hbox{\cmsss Z\kern-.4em Z}}
{\lower1.2pt\hbox{\cmsss Z\kern-.4em Z}}\else{\cmss Z\kern-.4em Z}\fi}
\def\1{{1\hskip -3pt {\rm l}}}




\lref\Inflation{
See, for example, 
A.~D.~Linde, 
{\it Particle Physics and Inflationary Cosmology}, 
(Harwood, Chur, Switzerland, 1990)\semi
A.~R.~Liddle and D.~H.~Lyth, 
{\it Cosmological Inflation and Large Scale Structure}, 
(Cambridge University Press, Cambridge, England, 2000)\semi
D.~H.~Lyth and A.~Riotto,
``Particle Physics Models of Inflation and the Cosmological Density
Perturbation,''
Phys.\ Rept.\  {\bf 314}, 1 (1999), 
{\tt hep-ph/9807278}.
}

\lref\sugra{
A.~B.~Goncharov and A.~D.~Linde,
  ``Chaotic Inflation in Supergravity,''
  Phys.\ Lett.\  B {\bf 139}, 27 (1984);
  ``A Simple Realization of the Inflationary Universe Scenario in SU(1,1)
  Supergravity,''
  Class.\ Quant.\ Grav.\  {\bf 1}, L75 (1984)\semi
H.~Murayama, H.~Suzuki, T.~Yanagida and J.~Yokoyama,
  ``Chaotic Inflation and Baryogenesis in Supergravity,''
  Phys.\ Rev.\  D {\bf 50}, 2356 (1994), 
  {\tt hep-ph/9311326}\semi
M.~Kawasaki, M.~Yamaguchi and T.~Yanagida,
  ``Natural Chaotic Inflation in Supergravity,''
  Phys.\ Rev.\ Lett.\  {\bf 85}, 3572 (2000), 
  {\tt hep-ph/0004243}\semi
J.~R.~Ellis, Z.~Lalak, S.~Pokorski and K.~Turzynski,
  ``The Price of WMAP Inflation in Supergravity,''
  JCAP {\bf 0610}, 005 (2006), 
  {\tt hep-th/0606133}.
}

\lref\Wess{
J.~Wess and J.~Bagger,
``{\it Supersymmetry and Supergravity},'' 
Princeton University Press (Second Edition, 1992).
}

\lref\Dvali{
  P.~Binetruy and G.~R.~Dvali,
  Phys.\ Lett.\  B {\bf 388}, 241 (1996)
  [arXiv:hep-ph/9606342].
}

\lref\FIterm{
  P.~Fayet and J.~Iliopoulos,
  Phys.\ Lett.\  B {\bf 51}, 461 (1974).
}

\lref\KY{
  K.~Kadota and M.~Yamaguchi,
  ``D-term Chaotic Inflation in Supergravity,''
  Phys.\ Rev.\  D {\bf 76}, 103522 (2007), 
  {\tt arXiv:0706.2676 [hep-ph]}.
}


\Title{                                \vbox{\hbox{UT-07-40}
                                             } 
}
{\vbox{\centerline{
                    Chaotic D-Term Inflation
}
}}

\vskip .2in

\centerline{
                     Teruhiko Kawano  
}

\vskip .2in 


\centerline{\sl
               Department of Physics, University of Tokyo
}
\centerline{\sl
                     Hongo, Tokyo 113-0033, Japan
}

\vskip 3cm
\noindent

A simple model for chaotic inflation in supergravity is proposed. 
The model is ${\cal N}=1$ supersymmetric massive U(1) gauge theory 
via the Stuckelberg superfield and gives rise to D-term inflation with a 
quadratic term of inflaton in the potential. 
The Fayet-Iliopoulos field plays a role of the inflaton.

\Date{December, 2007}



Cosmological inflation is a very attractive idea to solve the horizon problem 
and the flatness problem in the early universe \Inflation, and it also explains 
the density fluctuations in the cosmic microwave background. 
Thus, it would be interesting to incorporate the idea into viable models of 
high energy physics. Since supersymmetry is one of the solutions to the 
hierarchy problem between the electoweak scale and the Planck scale, 
it is natural to consider the inflation scenario in supergravity. 

Among many inflation scenarios, chaotic inflation doesn't suffer from 
the initial condition problems and gives rise to a large tensor-to-scalar ratio, 
which would be observed by the Planck satellite observation. 
However, it doesn't seem an easy task to find 
chaotic inflation models in supergravity \sugra. 
This is because the F-term 
potential in supergravity have the exponential factor of 
the K\"ahler potential in general to prevent inflaton fields from getting 
the large fluctuation required by chaotic inflation. One thus is led to 
make use of the D-term potential to give rise to chaotic inflation. 

In fact, the paper \KY\ is the first to give a chaotic inflation model by 
using the D-term potential, where it was shown that the quartic term of an 
inflaton field could be obtained from the D-term potential. Therefore, 
it would be interesting to obtain the quadratic term of inflaton fields 
from the D-term potential. Since in general, the D-term potential provides 
the quartic term of charged fields transforming linearly under the gauge group, 
it seems difficult to obtain the quadratic term of an inflaton field. 
However, the Fayet-Iliopoulos parameter of a U(1) gauge group enters 
the D-term potential quadratically. Thus, if one can replace the 
Fayet-Iliopoulos parameter by a dynamical field, one can expect that 
the field could play a role of an inflaton field. In this paper, 
it will be shown that this is indeed the case. 

Let us consider ${\cal N}=1$ supersymmetric massive U(1) gauge theory 
with the Stuckelburg superfield $S$ given in terms of component fields 
by
\eqn\stuckelburg{
S={1\over\sqrt{2}}\left(\rho+i\zeta\right)+\sqrt{2}\theta\psi+\theta^2F,
}
where $\rho$ and $\zeta$ are real scalar fields and play roles 
as the Stuckelburg field and the Fayet-Iliopoulos field, 
respectively, as will be seen. 
The U(1) gauge vector multiplet is given in the Wess-Zumino gauge by
\eqn\vector{
V=-i\theta\sigma^{\mu}\bar\theta v_{\mu}-i{\bar\theta}^2\theta\lambda
+i\theta^2\bar\theta\bar\lambda+\hf\theta^2{\bar\theta}^2D.
}
The Lagrangian density $L$ is given by 
\eqn\action{
L=-\hf\int d^2\theta d^2\bar\theta\left(S-\bar{S}+i\sqrt{2}M_pV\right)^2
+{\rm Re}\left[\int d^2\theta{1\over2e^2}W^{\alpha}W_{\alpha}\right],
}
where $M_p$ is a mass scale and the gaugino superfield $W_{\alpha}$ 
is given by 
\eqn\gaugino{
W_{\alpha}=-{1\over4}{\bar{D}}^2D_{\alpha}V.
}
The action is invariant under the U(1) gauge transformation 
\eqn\gauge{
V \to V+i\left(\Lambda-\bar\Lambda\right), \quad
S \to S+\sqrt{2}M_p\Lambda, \quad
\bar{S} \to \bar{S}+\sqrt{2}M_p\bar\Lambda.
}
Note here that the non-linear transformation of the chiral superfield $S$ 
prohibits it from appearing in the gauge kinetic function. 
Rescaling the vector superfield $V$ to obtain the canonical kinetic term 
for it, in component fields one finds the Lagrangian density 
\eqn\model{\eqalign{
L=
&-\hf\left(\d_\mu\zeta\right)^2
-\hf\left(\d_\mu\rho-Mv_\mu\right)^2
-{1\over4}F_{\mu\nu}^2
-i\bar\psi\bar\sigma^\mu\d_\mu\psi
-i\bar\lambda\bar\sigma^\mu\d_\mu\lambda
\cr
&+M\zeta{D}
-M\psi\lambda-M\bar\psi\bar\lambda
+\left|F\right|^2+\hf{D}^2,
}}
where $M$ is defined as $M=eM_p$.
It is obvious from \model\ that the field $\rho$ plays a role of 
the Stuckelburg field and gives mass to the gauge field $v_\mu$. 
The field $\zeta$ plays a role of the Fayet-Iliopoulos parameter, 
but here it is a real scalar field. 
Changing the variable $MA_\mu=Mv_\mu-\d_\mu\rho$ and solving the 
equation of motions of the auxiliary fields $D$ and $F$, one obtains 
the Lagrangian density
\eqn\action{
L=-\hf\left(\d_\mu\zeta\right)^2
-\hf M^2\zeta^2
-{1\over4}F_{\mu\nu}^2
-\hf M^2A_\mu^2
-i\bar\psi\bar\sigma^\mu\d_\mu\psi
-i\bar\lambda\bar\sigma^\mu\d_\mu\lambda
-M\psi\lambda-M\bar\psi\bar\lambda.
}
This is the Lagrangian density of a free massive U(1) vector multiplet.
The Fayet-Iliopoulos field $\zeta$ plays a role of inflaton.
Therefore, one expects that the mass $M$ of the inflaton field $\zeta$ 
is of order $10^{13}\sim10^{14}$ GeV. Thus, if the mass scale $M_p$ 
is set to the reduced Planck scale $\sim 10^{18}$ GeV, 
the $U(1)$ gauge coupling may be of order $10^{-4}\sim10^{-5}$ to give rise 
to the observed density fluctuations. However, one can see that the reheating 
doesn't occur without any modification in the model. Since the model is very 
simple, we hope that a simple modification of it could lead to a full-fledged 
inflation model.


\vskip 0.3in
\centerline{{\bf Acknowledgements}}
\medskip
The author would like to thank Koichi Hamaguchi, Yuuki Shinbara, 
Taizan Watari, and Masahide Yamaguchi for helpful discussions and 
reading the manuscript. 
The work was supported in part by a Grant-in-Aid (\#19540268) 
from the MEXT of Japan.


\listrefs

\end